*Article*

# The 3D Direct Simulation Monte Carlo Study of Europa's Gas Plume

Wei-Ling Tseng[1], Ian-Lin Lai[1,2,*], Wing-Huen Ip[3,4], Hsiang-Wen Hsu[5] and Jong-Shinn Wu[6]

Affiliations:
1. Department of Earth Sciences, National Taiwan Normal University, Taipei, Taiwan
2. Physikalisches Institut, University of Bern, Bern, Switzerland
3. Institute of Astronomy, National Central University, Taoyuan, Taiwan
4. Space Science Institute, Macau University of Science and Technology, Macau, China
5. LASP, University of Colorado Boulder, CO, USA
6. Department of Mechanical Engineering, National Yang Ming Chiao Tung University, Hsinchu, Taiwan

*Corresponding author: Dr. Ian-Lin Lai; ianlai@ntnu.edu.tw

**Abstract:** Europa has been spotted to have water outgassing activities by the space and ground-based telescopes as well as reanalysis of the Galileo data (Roth et al. 2014; Sparks et al. 2016, 2017; Paganini et al. 2020; Jia et al. 2018; Arnold et al. 2019). We adopt a 3D DSMC model to investigate the observed plume characteristics of Europa assuming supersonic expansion originated from the undersurface vent. With a parametric study of the total gas production rate and initial gas bulk velocity, the gas number density, temperature and velocity information of the outgassing plumes from the various case studies are derived. Our results show that the plume gases experience acceleration through mutual collisions and adiabatic cooling when exiting from the surface. The central part of the plume with the relatively large gas production rates (of $10^{29}$ and $10^{30}$ $H_2O$ $s^{-1}$) is found to sustain thermal equilibrium and nearly continuum condition. Column density maps integrated along two different viewing angles are presented to demonstrate the importance of the projection effect on remote sensing diagnostics. Finally, the density profiles at different altitudes are provided to prepare for observations of Europa's plumes including the upcoming spacecraft missions such as JUICE and Europa Clipper.

## 1. Introduction

Europa is an icy moon of Jupiter, likely to host a subsurface ocean as derived from a diversity of the surface geological features and the magnetic field measurements made by the Voyager and Galileo missions (e.g. Anderson et al. 1998; Carr et al. 1998; Kivelson et al. 2000; Greenberg, 2002). Recently, outgassing plumes have been inferred to occur on Europa, likely being originated from its subsurface liquid water reservoir(s). These observations include (1) the local density inhomogeneity of the atomic O and H aurora emissions observed by the Hubble Space Telescope (HST), whose emission intensities are consistent with electron-impact dissociation of water molecules (Roth et al. 2014); (2) the HST observation of absorption features shown in the UV continuum of Jupiter during Europa's transits (Sparks et al. 2016, 2017; Giono et al. 2020); and (3) the water molecule observed in the infrared wavelength by the Keck Observatory (Paganini et al. 2020). In addition, reanalysis of the Galileo data including the magnetic field and plasma measurements also inferred that Europa had eruptions (Jia et al. 2018; Arnold et al. 2019).

All above reports suggested that the plumes originated from various latitudinal regions in the southern hemisphere, except for no latitudinal information of the Keck result due to its observational configuration. Neither the exact source location nor the venting mechanism (i.e. plume dynamics) is explicitly disclosed. The gas thermal velocities were required to be at least 0.5 km s$^{-1}$ and 0.7 km s$^{-1}$ to arrive at the altitude of the observed peaks of ~100 km (Sparks et al. 2017) and ~200 km (Roth et al. 2014), respectively, assuming plume material follows ballistic motion. The water column densities of ~10$^{20}$ m$^{-2}$ and 1.8x10$^{21}$ m$^{-2}$ were derived by Roth et al. (2014) and Sparks et al. (2017), respectively, while the Keck measurements of Paganini et al. (2020) suggested 1.4x10$^{19}$ H$_2$O m$^{-2}$. This could suggest that a mixture of various outgassing activities occurred in random places on Europa. The durations of Europa's plume eruptions are unknown, either, which are only provided with the integrated observation time of several hours (e.g. ~2.5 hrs from Paganini et al. 2020; ~ 1hr from Sparks et al. 2017; ~7 hrs from Roth et al. 2014). The signatures of Europa's plumes inferred by the Galileo data, i.e. the E12 flyby at the closest-approach altitude of ~200 km (Jia et al. 2018) and the E26 flyby at the closest-approach altitude of ~350 km (Arnold et al. 2019), were roughly consistent with the small-scale plume characteristics suggested by Roth et al. (2014). The in-situ measurements of Galileo could give better constraints on the source regions which were both in the southern hemisphere and trailing side. However, existing observations also indicated that the Europa's outgassing activity may not be similar to Enceladus' persistent plumes but were sporadic and/or intermittent events (Roth et al. 2016; Paganini et al. 2020; Sparks et al. 2019).

In addition to recent observations, modeling efforts have been made to investigate Europa's plume morphology and its consequences. For example, some surface features on Europa could be associated with its plume emplacements via cryovolcanism (Fagents et al. 2000; Quick et al. 2013). Using a ballistic trajectory model, Quick et al. (2013) found that the small-scale plumes with a low eruption velocity (< 0.3 km s$^{-1}$) could account for the dark material deposited around the ridges, lenticulae and lineae shown in the Galileo images. Quick and Hedman (2020) examined the morphological and spectral signatures of the possible plume deposits, and used the latest HST observations to constrain the eruption velocity to be 0.03 – 1.0 km s$^{-1}$. Southworth et al. (2015) modeled Europa's dust plume morphology and studied its surface deposition effect assuming that the driving mechanism is similar to that of Enceladus (Schmidt et al. 2008). When investigating Europa's global exosphere mainly generated from sputtering and thermal adsorption, Teolis et al. (2017) also added an assumed plume source with a production rate of 10$^{28}$ H$_2$O s$^{-1}$ in the south polar region. Their modeling results showed that a significant gas density enhancement could exist near the plume source region (i.e. lifetime on the order of magnitude of 10$^5$ sec). Vorburger and Wurz (2021) used a Monte-Carlo model with a ballistic approach to simulate the plume profiles from three different source mechanisms, and found out that the source can not be determined by comparing the simulated atomic H and O emissions with the existing HST observations (Roth et al., 2014; 2014b).

In this work, a DSMC model is developed to investigate the plume characteristics of Europa. The Direct Simulation Monte Carlo (DSMC) method proposed by Bird (1994) is suitable for studying the rarefied gas flow by solving the Boltzmann equation. For example, the physical processes and structures of the planetary jets and plumes can be adequately described by the DSMC model with a broad flow regime ranging from close to semi-continuum near surface, to collisionless condition at high altitudes (e.g., Zakharov et al. 2009; Tenishev et al. 2011; Lai et al. 2015; 2019; Yeoh et al. 2015; 2017; Goldstein et al. 2018; Zhang et al. 2004). Recently, Berg et al. (2016) used a DSMC model to examine Europa plumes with a primary variable of the Mach number which dominates the plume expansion. Their Mach number is determined by the throat-to-vent area ratio assuming that the supersonic expansion is originated from a subsurface vent similar to Enceladus' condition (Schmidt et al. 2008). Understanding the plume dynamics above surface, therefore, can be used to probe its subsurface vent properties. In addition, a canopy shock layer can be formed in the top of plume with a high flow mass rate of ~1,000 kg s$^{-1}$ which would limit the overall plume height (Berg et al. 2016).

We conduct a systematic study about the effects of total gas production rate and initial gas bulk velocity with an assumption of supersonic eruption from subsurface. The detail of our DSMC model and justifications of the model inputs are presented in SECTION 2. SECTION 3 shows the simulated plume results of Knudsen number, gas temperature, number density and velocity distributions. The projection effect on the plume morphology integrated along two different viewing angles is also discussed. The density vs. horizontal-distance profiles of the simulated plumes at different altitudes are provided for comparisons with any possible future observation. Comparisons with the existing plume observations and the important consequences of Europa's outgassing activity are discussed at the end of SECTION 4. Finally, a summary of our work is given in SECTION 5. The modeling results of both assumed gas production rate and gas bulk velocity (such as plume height, gas temperature, velocity and density distribution) can be used to improve scientific interpretations of the space and ground-based telescope observations of Europa's plumes. It is also helpful for the designs of the in-situ measurements of future spacecraft missions such as Juno, JUICE and Europa Clipper, to characterize further information about the plume source region on Europa.

## 2. Materials and Methods

### 2.1. General Description

The DSMC method is a particle-base method to simulate gas flow with a large number of particles. This has been proposed by Bird (1994), which solves the Boltzmann equation for all regions of rarefaction. The Knudsen number (Kn) indicates the rarefaction of gas flow and is defined as $Kn = L/\lambda$, where L is the local characteristic length and $\lambda$ is the local mean free path. If $Kn > 0.001$, the continuum assumption with a conventional no-slip boundary condition breaks down (Cercignani, 1998). The PDSC++ code used in the present study is a 3D parallelized DSMC code including several features: unstructured mesh, variable time-step scheme, domain re-decomposition, automatic steady state detection scheme (Wu et al. 2004; Su et al. 2010; Cave et al. 2008; Lo et al. 2015). It also has been successfully applied to the modeling of cometary outgassing (Finklenburg et al. 2014; Lai et al. 2017; Liao et al. 2018; Marschall et al. 2016).

To calculate the plume of Europa, we build an unstructured grid with 3,987,633 tetrahedron cells. The cell size is set to be smaller than the local mean free path of the molecules. The size of the simulation box is about 12 Europa's radii. The initial gas temperature is assumed to be in thermodynamic equilibrium (i.e. kinetic temperature equal to rotational temperature) when exiting from the surface boundary condition. A general DSMC calculation can be described as following:

1. Set initial state and read system data: Before calculation, the DSMC program has to set up a grid for the calculation, and input the initial condition and boundary condition.

2. Move all particles: In the calculation of each time step, the simulation gas particles travel a distance with their velocities. The particles which leave the simulation domain will be removed.

3. Introduce new particles: After moving all the particles, new particles will be added into the simulation domain from the boundary of the source region.

4. Sort particles and calculate collisions: In this phase, all particles are indexed in the simulation domain and identify the collision pairs depending on the local number density and calculate the new velocities after collision. In this work, we use the variable soft sphere (VSS) model for the collision. In addition, collisions also allow for energy transfers between the translational and internal (i.e. rotational) mode.

5. Sample the flow and check the steady-state flow: In each cell, the macroscopic properties of flow are sampled from the microscopic properties of each particle. When the flow reaches the steady state by checking the convergence of the total particle number, velocity and temperature, the flow field will be stored. The final result of the gas flow will be calculated by taking the average of the sufficient sampling.

### 2.2. A Parametric Study of the Gas Production Rate and Gas Bulk Velocity

In this work, we examine the physical processes and structures of supersonic gas plumes of Europa. Neither grain formation due to condensation nor collision with grains is taken into account. The effects of grains on the planetary plume morphology are found to be negligible (e.g. Yeoh et al. (2015)). The only loss process of the water molecule considered here is to re-impact the surface meaning that the modeled plume reaches a steady state not much longer than the particle flight time (i.e. on the order of several thousand seconds). By the same token, the water loss to photolytic and electron-impact reactions is not included because of the much longer reaction timescales (i.e. > ~$10^5$ seconds). Coriolis force and centrifugal force are found negligible when the particle flight time is much shorter than Europa's rotation period. A circular source area is assumed with a radius of 20 km near a pole on Europa. Noted that it is not a representative size of a realistic vent on Europa because there is no actual measurement and/or solid evidence of the undersurface vent size on Europa. Regardless of the meter-sized vent on Enceladus suggested from Schmidt et al. (2008), the vent size on Europa can be expected to be broader due to its larger mass (e.g. Fagents 2003). A large modeled vent size can be seen as a collective effect of many small vents located in a limited region, and we focus only on the macroscopic phenomenon assuming that these smaller plumes do not have strong interactions to significantly modify the overall plume morphology. A similar approach has also been used in Zhang et al. (2004) to study Io's volcanic plumes with a "virtual" vent radius of 16 km in a DSMC model. In addition, the observable outgassing properties such as plume height and width are found to change only in a range of 10-20% when modifying the modeled parameters such as the vent size. In addition, the spatial resolution (~200 km) and sensitivity in the HST images (e.g. Roth et al. 2004) are insufficient to distinguish the collisional part and non-collisional part of observed plumes. Our approach is thus adequate to use the observed plume morphology (i.e., plume height and width) to investigate the gas eruption properties near surface. Although it is a compromising strategy between the computational time and spatial resolution, the modeling results is a useful first step for comparing with remote sensing observations.

As discussed above, the existing observations did not give a clear and consistent picture of Europa's plume eruption. Besides, it has also been observed that Enceladus' outgassing activities could be a mixture of various gas emission rates from different source locations, narrow & high-altitude jets, and diffusive & low-altitude plumes (e.g. Spitale and Porco, 2007; Porco et al. 2014). Therefore, we study Europa's plume activity with a large parametric space mainly relying on what we learned from Enceladus. In the simulations, the initial gas thermal temperature of Europa's plume is assumed to be 180 K based on the source temperature of Enceladus' plume gas (Spencer et al. 2006) considering the lack of direct constraints of Europa's plumes. The previously published temperatures of Europa's gas plumes were indirectly derived from the altitudes of the observed signals assuming plume material following ballistic trajectories with their gas thermal velocities could reach (e.g. >230K from Roth et al. 2014). Note that the plume peak altitudes derived from observations may not represent the actual plume top due to projection effect (i.e. the line-of-sight effect and the unknown source region), as will be discussed in Section 3.3. The supersonic & high-altitude jets (i.e. high Mach number), and the diffusive & low-altitude plumes (i.e. low Mach number) observed in Enceladus, on the other hand, suggested both medium- and high-velocity ejections (e.g. Spitale and Porco, 2007; Porco et al. 2014; Dong et al., 2011; Hansen et al. 2008 & 2011). Along this line, a systematic study of the gas bulk velocities from subsonic (0.35 km s$^{-1}$) to supersonic condition at 0.5 km s$^{-1}$, 0.75 km s$^{-1}$ and 1.0 km s$^{-1}$ with averaged Mach numbers between 1-2 are carried out in this work.

Finally, three water gas production rates of $10^{28}$ s$^{-1}$, $10^{29}$ s$^{-1}$, $10^{30}$ s$^{-1}$ are considered in the simulations. These values appear to be much higher than the Enceladus plume source rates (i.e. in the order of $10^{28}$ s$^{-1}$), with the middle case being comparable to the values derived by Roth et al. (2014) (~2.3x$10^{29}$ H$_2$O s$^{-1}$) and Paganini et al. (2020) (~8x$10^{28}$ H$_2$O s$^{-1}$). It is also found that the Europa plume mass was ~100 times that of Enceladus when Hansen et al. (2019) compared the Enceladus plume material measured by Cassini UVIS (Ultraviolet Imaging Spectrograph Subsystem) with the absorption features of Europa's plumes seen in the HST images (Sparks et al. 2017). The initial model inputs for the 6 case studies are summarized in **Table 1**. While cases 1-3 are simulated with the same gas bulk velocity of 0.5 km s$^{-1}$ to study the effect of varying gas production rate on the plume morphology, the influence of initial gas bulk velocity is examined by cases 4-6 with the same Q = $10^{29}$ $H_2O$ s$^{-1}$. A special case 7 is included to investigate the possible stealth plume activity under the current detection limits (i.e. a very low gas production rate of $10^{26}$ $H_2O$ s$^{-1}$) which could be discovered by the upcoming space missions (discussed in Section 4).

**Table 1. The initial conditions in the case studies with a circular source area of a radius of 20 km.** *A special case 7 is to simulate a possible stealth plume with a circular source area of a diameter of 10 km.

|  | Case 1 | Case 2 | Case 3 | Case 4 | Case 5 | Case 6 | Case 7* |
|---|---|---|---|---|---|---|---|
| Gas Production Rate ($H_2O$ s$^{-1}$) | 1x$10^{30}$ | 1x$10^{29}$ | 1x$10^{28}$ | 1x$10^{29}$ | 1x$10^{29}$ | 1x$10^{29}$ | 1x$10^{26}$ |
| Gas Bulk Velocity (km s$^{-1}$) | 0.5 | 0.5 | 0.5 | 0.35 | 0.75 | 1.0 | 0.35 |
| Gas Temperature (K) | 180 | 180 | 180 | 180 | 180 | 180 | 180 |
| Mach Number | 1.1 | 1.1 | 1.1 | 0.8 | 1.6 | 2.2 | 0.8 |

*A special case 7 is to simulate a possible stealth plume with a circular source area of a radius of 5 km.

## 3. Results

Here we show the simulation results of 6 case studies to investigate the physical structure of Europa's outgassing plume expanding from surface to high altitude. Both effects of the total gas production rate and gas bulk velocity on the plume morphology and dynamical process will be discussed.

### 3.1. Knudsen Number, Gas Temperature and Thermal Equilibrium

**Figure 1 (column a)** shows the modeled local Knudsen number, Kn, along the plume symmetric plane. It is seen that the Kn varies by several orders of magnitudes across the plume. The gas flow in the low-altitude and central region of the plume shows a nearly continuum condition (Kn ≤ 0.01) where frequent gas collisions occur. Then, it expands to the collisionless condition (Kn ≥ 1) at a high altitude (i.e. transition region). The transition altitude where the flow starts to deviate from continuum is highest (~0.5 Europa radius above surface) with the highest gas bulk velocity of 1.0 km s$^{-1}$ (Figure 1: case 6 (a)). In addition to the effect of initial gas bulk velocity, the transition height slightly decreases with decreasing gas production rate, particularly for $10^{28}$ s$^{-1}$ due to insufficient gas collisions (see Figure 1: case 3 (a)).

The degree of local thermal equilibrium represented by γ is shown in **Figure 1 (column b)**. γ is defined to be "the absolute value of ($T_{trans}/T_{rot} - 1$)" where $T_{trans}$ and $T_{rot}$ are the gas translational temperature and rotational temperature, respectively. When γ is close to zero, the gas flow approaches thermal equilibrium (i.e. the translational temperature equal to the rotational temperature), which mostly occurs in the dense part of the plume along its central line and its upper layer, except in the case 3 with the lowest gas production rate of $10^{28}$ s$^{-1}$. At a very top of the plume, γ becomes larger and $T_{trans}$ deviates from $T_{rot}$. The energy in each mode (i.e. translational and rotational temperature) may remain "locked" without sufficient gas collisions allowing for energy transfer in different modes. By the same token, the translational and rotational temperatures differ from one another along the lateral edges of the plumes (i.e. red color in the bowtie-shaped region as discussed below) where the gas collision frequency is too low to maintain thermal equilibrium. Similar features are also reflected in the Kn plots. The chaotic regions outside of the major plume tops (i.e. the red-color regions) are the results of very few particles scattered by gas collisions. As expected, the degree of thermal equilibrium depends on the collision frequency, which is also controlled by the gas production rate (see cases 1-3 in Figure 1(b) for comparisons). The distribution of the degree of thermal equilibrium is generally consistent with the one of Kn where a smaller Kn indicates a condition closer to thermal equilibrium (i.e. smaller γ). **Figure 1 (column c)** presents the gas translational temperature distributions. The gas translational temperature along the center line of the plume rapidly drops from the initial 180 K to below 100 K within an altitude of ~100 km. It is due to plume expansion that causes adiabatic cooling primarily in the low altitude region**.** The statistical noise above the plume tops (Figure 1: case 6 (c)) are the results of low counting statistics in this region, which is also seen in the lateral sides of the plume bottom (i.e. the high temperature region) where too few particles were recorded.

*3.2. Gas Number Density and Velocity Distributions*

The number density distributions of the simulated plumes are shown in **Figure 2 (column a)**. Because of expansion, the gas number density drops rapidly with the increasing altitude. With the same initial velocity, the gas number density generally increases with the larger gas production rate (case 1-3(a) in Figure 2). There are also some distinct features such as the bow-shape structure (i.e. canopy) at the plume top and the "bowtie" pattern in the lateral sides of the plume in the low-altitude region. Both features are seen in Figure 1 as well. The enhanced density shown in the "bow" shape at the plume top is due to the sum of two particle populations (i.e. upward flow and downward flow) which approach zero velocities. This feature is mostly obvious in the plume produced with the high gas production rate of $10^{30}$ s$^{-1}$ and $10^{29}$ s$^{-1}$. The importance of canopy shock layer on the plume morphology has also been discussed in Berg et al. (2016). The "bowtie" pattern represents the voids in the lateral sides of the plume where very few particles travel through. With the same gas production rate, the effect of initial gas ejection velocity on the plume morphology is presented in Figure 2 for case 2, 4-6 for Q=$10^{29}$ s$^{-1}$ respectively, which clearly show that the reducing gas ejection velocity leads to a smaller and much denser plume.

The gas velocity distribution of the plume is shown in **Figure 2 (column b)**. Overall, the gas velocity decreases with increasing altitude because of Europa's large gravity. However, when examining the velocity evolution below the altitude of ~ 100 km, it is found that the gas velocity appears to be slightly larger than the initial gas bulk velocity (i.e. see the obvious case 5 & 6; other cases are not clearly shown due to the color scale). This can be attributed to the acceleration due to gas collision and adiabatic cooling during expansion. The averaged gas flow velocity along the central line of plume can be accelerated to ~0.6, 0.65, 0.85 and 1.05 km s$^{-1}$ from its initial gas bulk velocity of 0.35, 0.5, 0.75 and 1.0 km s$^{-1}$, respectively. Because of this acceleration effect, therefore, the plume top from the DSMC modeling would be much higher than the one with purely ballistic motion (discussed in Section 3.3 and 3.4). The degrees of lateral expansion resulting from gas collisions of various gas production rates are also shown in case 1-3 in Figure 2. Following the streamlines, the plume particle speed also increases when returning to surface. The plume gas molecules mostly fall back to Europa's surface since the initial ejection velocity is lower than Europa's escape velocity of ~ 2.0 km s$^{-1}$. Figure 3 (a) shows the plume profiles of number density as a function of distance at four altitudes of 50, 100, 200 and 400 km (i.e. a horizontal cut through the plume). It is clearly shown that the plume width (e.g. the FWHM (Full Width Half Maximum) of the 50-km profile) is mainly controlled by the Mach number, while the gas collision (determined by the gas production rate) plays a minor role. The distinct density profiles presented in Figure 3(a) can be used to compare with future in-situ neutral particle detection to examine the plume properties of Europa outgassing activities.

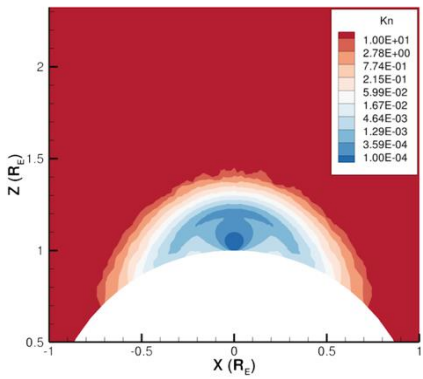
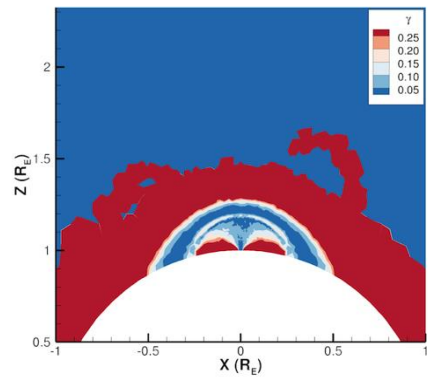
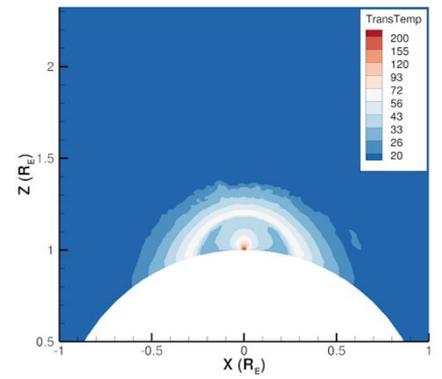
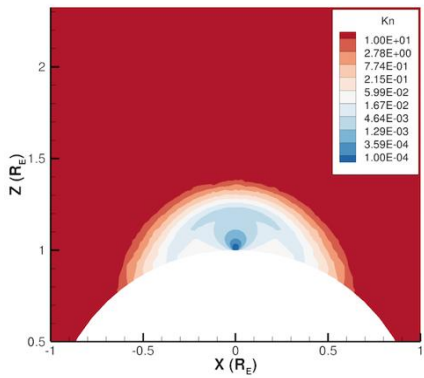
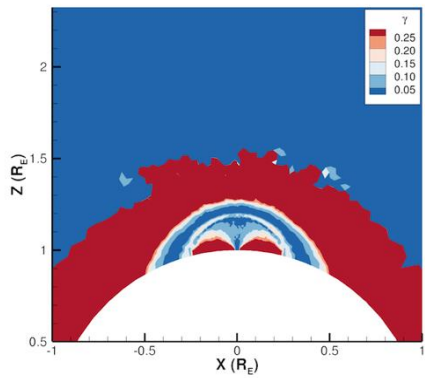
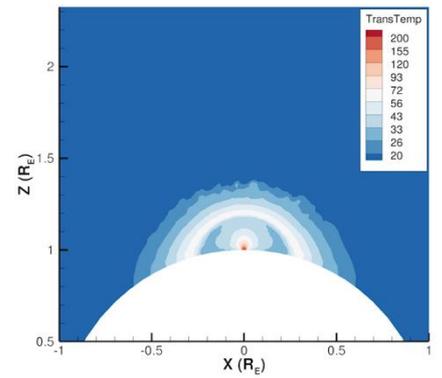
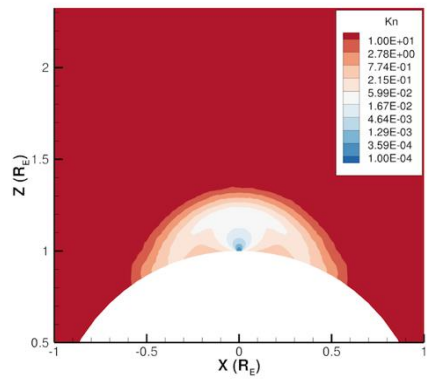
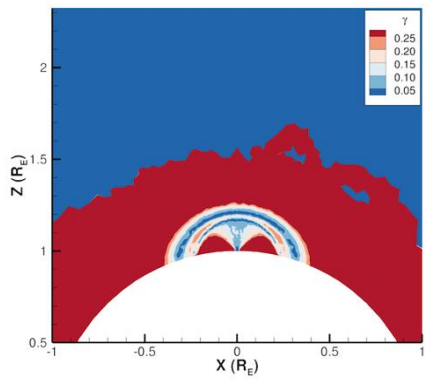
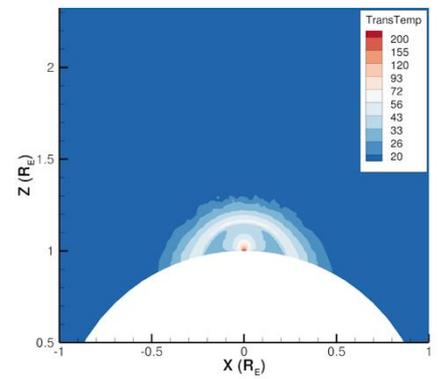

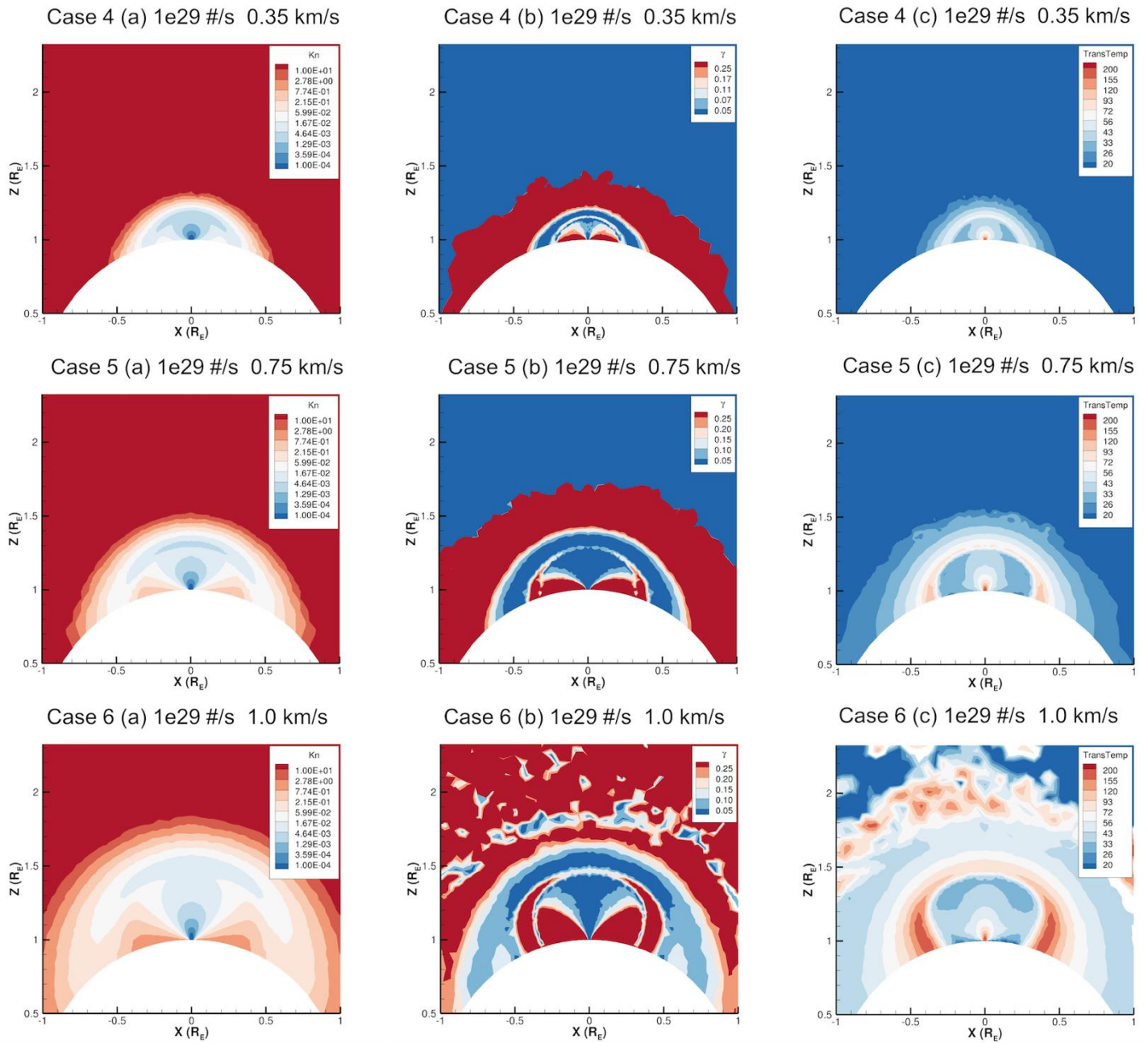

**Figure 1. Column a (Left)**: Results of local Knudsen number, $K_n$ (shown in contours with a color scale). When $K_n > \sim 0.01$, the gas flow starts to deviate from continuum. **Column b (Middle)**: Degree of thermal equilibrium represented by $\gamma$ (shown in contours with a color scale of $\gamma$). $\gamma$ is defined to be "the absolute value of ($T_{trans}/T_{rot}$) minus 1". $\gamma$ closer to 0 indicates that the gas flow approaches thermal equilibrium. **Column c (Right):** Gas translational temperature distribution (shown in contours with a color scale in a unit of K). Both axes are arbitrary and presented in the unit of Europa radius. Poor statistics is shown above the major plume tops (i.e. red color). This is also shown in the lateral sides of the plume bottoms where too few particles were counted.

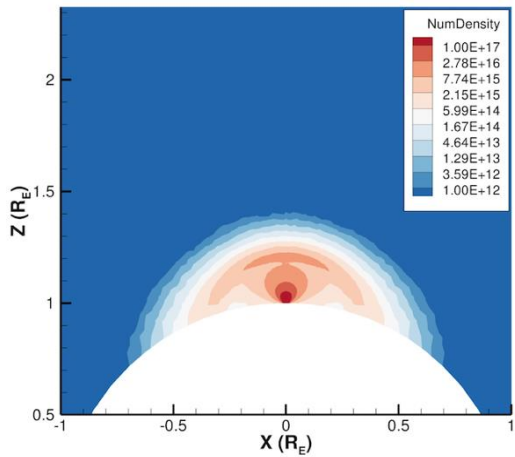
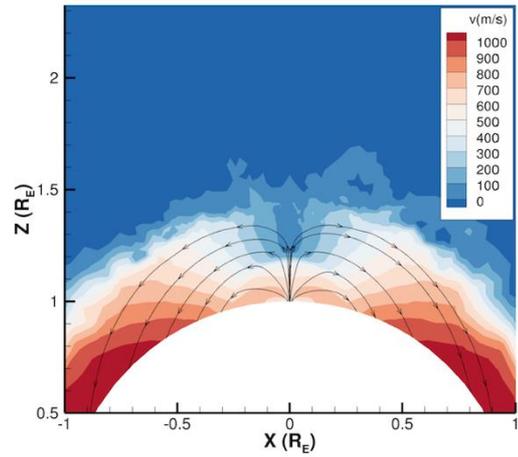
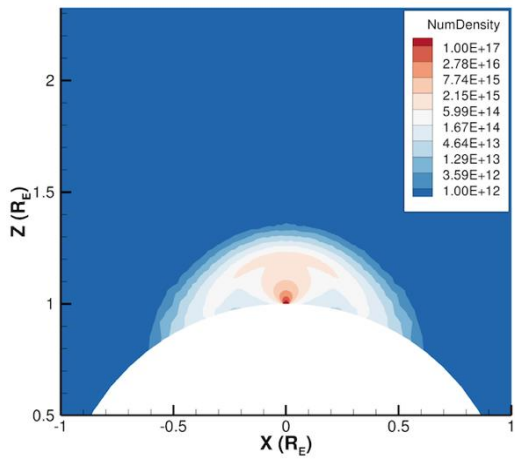
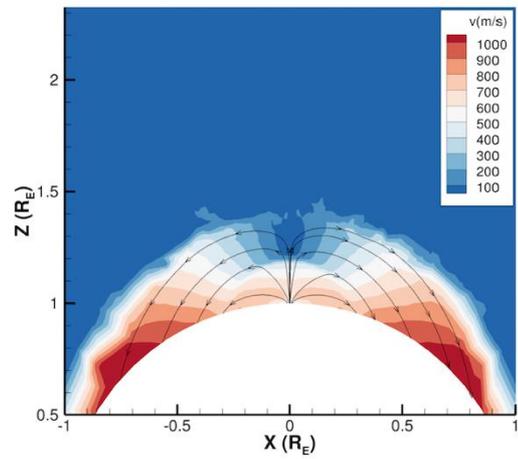
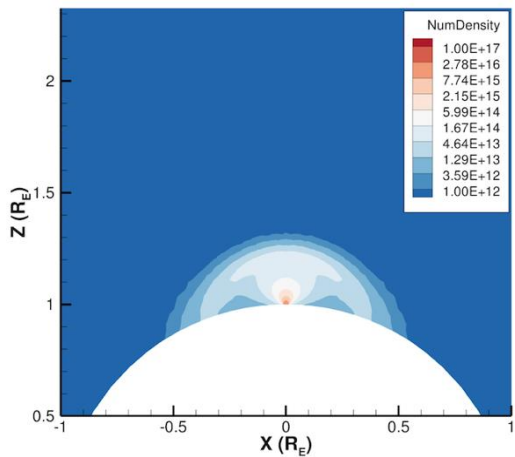
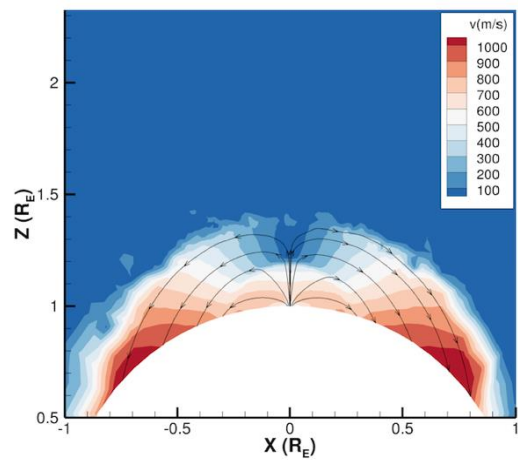

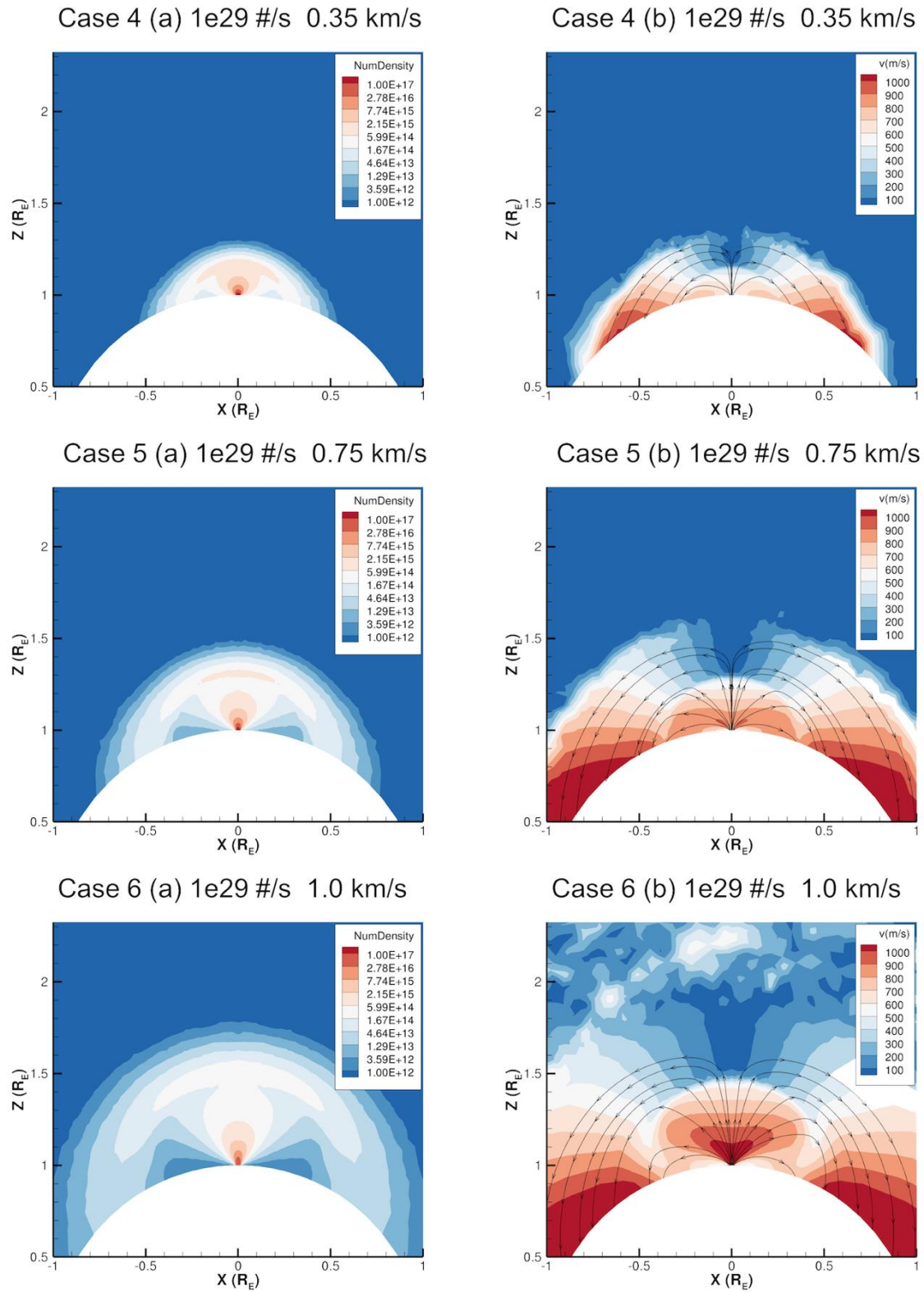

**Figure 2. Column a (Left):** Number density distribution (shown in contours with a color scale in a unit of # m$^{-3}$). **Column b (Right):** Gas velocity distribution (shown in contour with a color scale in a unit of m s$^{-1}$). Gas flow streamlines are presented too.

Both axes are arbitrary and presented in the unit of Europa radius.

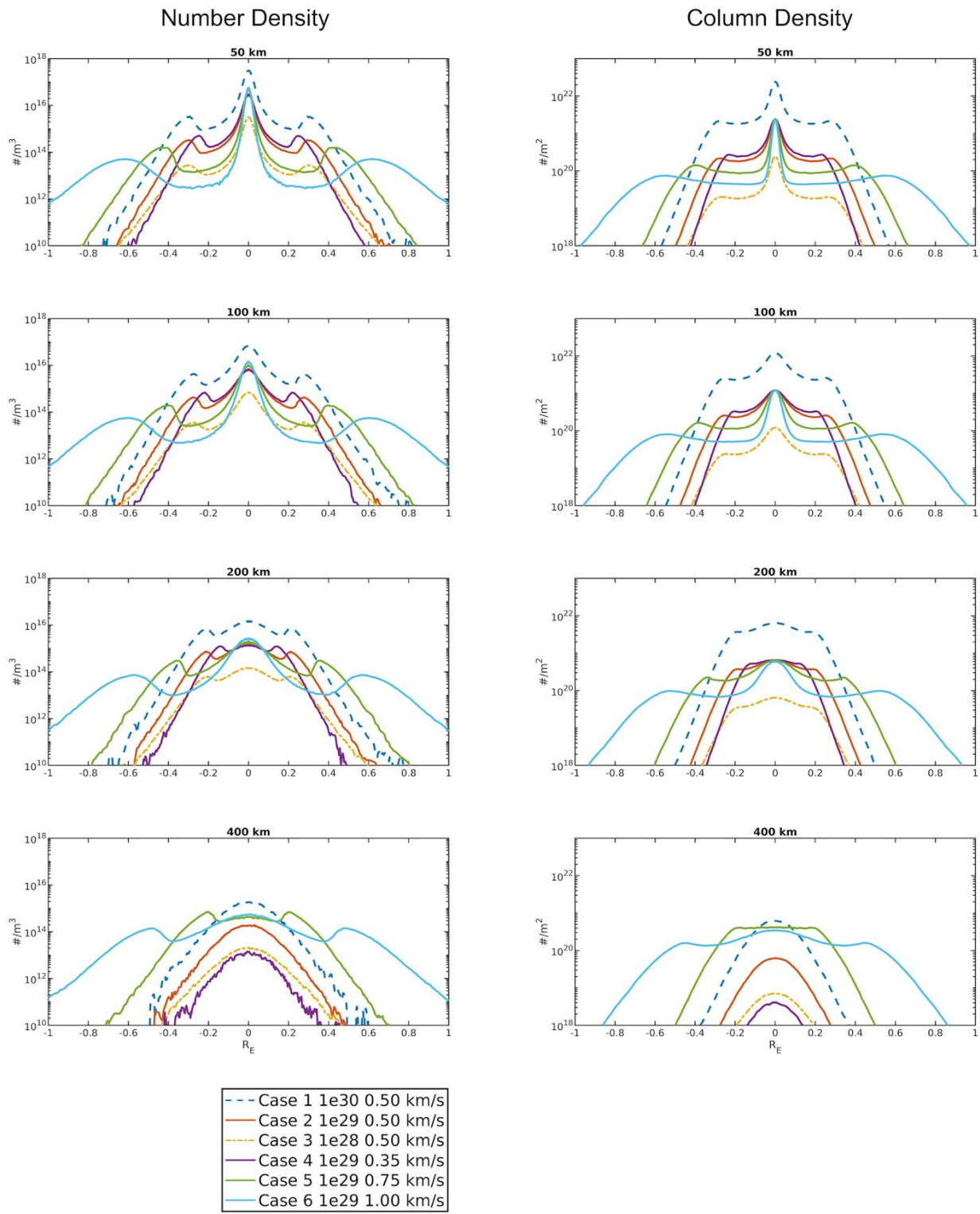

**Figure 3. Column a (Left):** The plume profiles of number density (#/m³) as a function of horizontal distance at four altitudes of 50, 100, 200 and 400 km. **Column b (Right):** The plume profiles of column density (integrated along a line of sight of 90 degrees) as a function of horizontal distance at four altitudes of 50, 100, 200 and 400 km.

The profiles are given in a mocked spacecraft trajectory flying across the plume horizontally (i.e. parallel to the x-axis in Figure 2(a) and 4(a)) at different altitudes. The x-axis is presented in the unit of Europa radius.

*3.3. Column Density*

The simulated plume morphologies with the integrated column densities along two different viewing angles of 90° (i.e. perpendicular to the plume symmetric axis) and 45° are presented in **Figure 4**. Figure 4 (column a) (along a line of sight of 90°) show that the height of the dense plume (i.e. red/pink color) is mainly determined by the initial gas ejection speed, and the plume heights are ~250 km, ~350 km, ~500 km and ~800 km for 0.35 km s$^{-1}$, 0.5 km s$^{-1}$, 0.75 km s$^{-1}$ and 1.0 km s$^{-1}$, respectively. As discussed above, the plume height appears to be much higher than the one computed with purely ballistic motion as a result of acceleration due to gas collisions in the low altitude region. For example, the ballistic particle with an initial ejection speed of 0.75 km s$^{-1}$ can reach a maximum height of only ~210 km, less than half of the modeled height. On the other hand, when viewing from 45° as shown in Figure 4 (column b), the plume extends down to the surface and covers a part of Europa's disk. It is evident that the observed plume morphology strongly depends on the viewing geometry. This also shows that it is not an easy task, without any other supplementary information, to trace back a plume seen in one 2D projected image to its exact source region. The column density profiles (integrated along a line of sight of 90°) as a function of horizontal distance to the plume symmetric axis at four altitudes of 50, 100, 200 and 400 km are shown in Figure 3 (b). The maximum column densities at an altitude of 50 km simulated with the water production rates of $10^{28}$ s$^{-1}$, $10^{29}$ s$^{-1}$ and $10^{30}$ s$^{-1}$ are in the order of $10^{20}$ m$^{-2}$, $10^{21}$ m$^{-2}$ and $10^{22}$ m$^{-2}$, respectively.

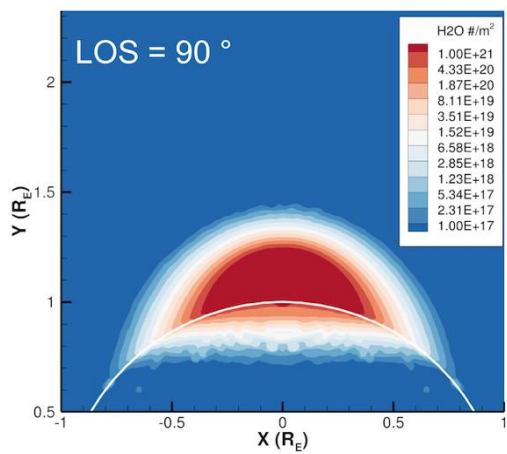
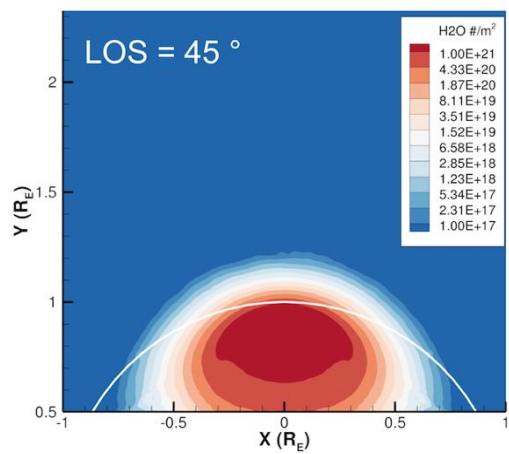
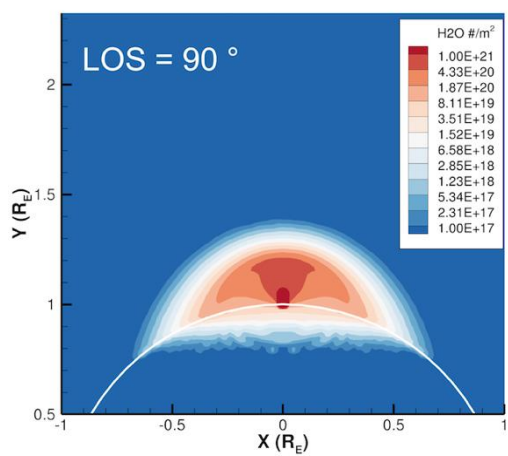
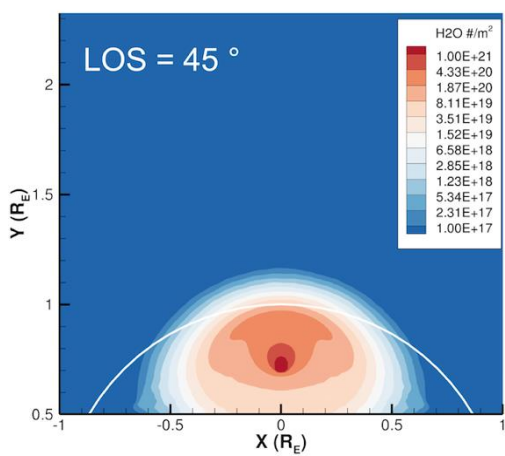
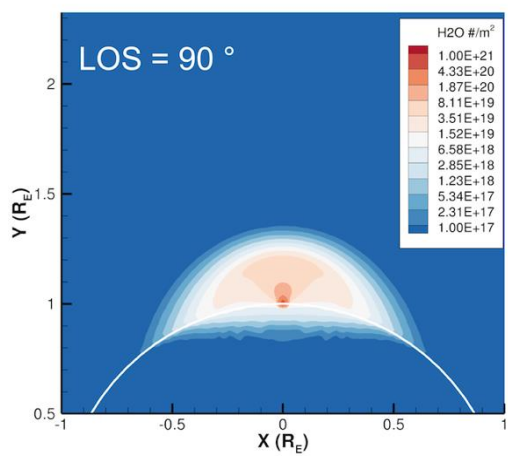
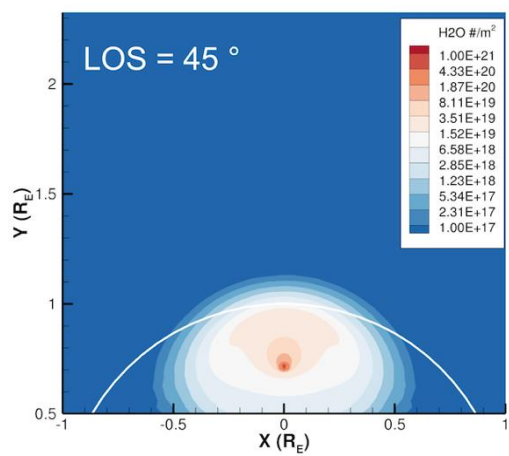

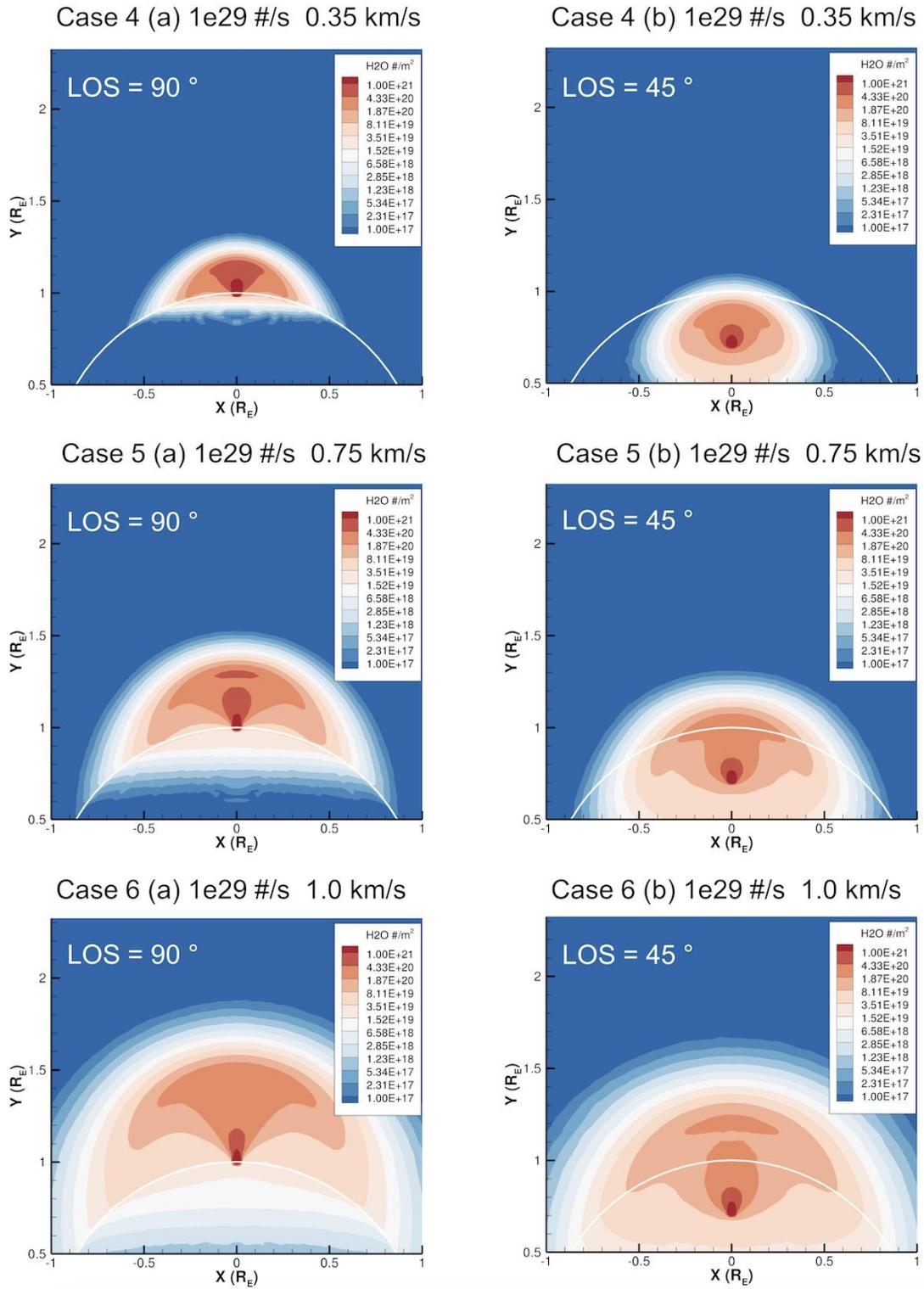

**Figure 4. Column a (Left):** Column density integrated along a line of sight of 90 degrees from Europa's pole (i.e. perpendicular to the paper). **Column b (Right):** Column density integrated along a line of sight of 45 degrees from Europa's pole (i.e. toward the reader).

The contour is shown in a color scale in a unit of # m$^{-2}$. Both axes are arbitrary and presented in the unit of Europa radius.

## 4. Discussions

Compared with the HST atomic emission data (Roth et al., 2014) and the Galileo E12 measurements (Jia et al. 2018) obtained at an altitude of ~200 km, our simulation cases of gas production rate Q = 1.0x10$^{29}$ s$^{-1}$ (i.e. see case 2, 4, 5 and 6 in Figure 3(b)) produce the averaged column densities in the order of 10$^{20}$ m$^{-2}$ which are consistent with the derived column densities of the HST and Galileo observations when taking into account of the spatial resolution limited by the observations (i.e. ~ 100 km in Roth et al. 2014). In addition, the peaked column densities measured by Sparks et al. (2017) (~1.8x10$^{21}$ m$^{-2}$) and by Paganini et al. (2020) (~1.8x10$^{19}$ m$^{-2}$) are closer to the simulated case of the production rate of 10$^{30}$ s$^{-1}$ (case 1) and 10$^{28}$ s$^{-1}$ (case 3), respectively. The simulated plume tops of the initial gas bulk velocity of 0.35 km s$^{-1}$ and 0.5 km s$^{-1}$ reach an altitude of ~250 km and ~350 km which are also in agreement with of the HST and Galileo measurements (Roth et al. 2014; Jia et al. 2018; Arnold et al. 2019). These initial gas velocities are smaller than the ones derived by the HST observations in which the authors assumed the ballistic particle motion (Roth et al. 2014; Sparks et al. 2017). As discussed above, the plume morphologies (i.e. the height and width) simulated from the DSMC modeling are much different from purely ballistic motion due to gas collisions occurred in the low altitude region. Moreover, it is difficult to interpret the observed 2D projection images because the combined effect of the viewing angle and the gas velocity is degenerated, meaning that the solution may not be unique if the plume source region is not known. The in-situ measurements of the upcoming ESA JUICE and NASA Europa Clipper missions with finer spatial resolution are expected to reveal the exact source regions and the detailed plume structures such as the gas density, velocity and temperature distributions. These parameters are important tools to probe its undersurface vent properties and investigate its surface deposition effect to be discussed below. Finally, a special case 7 (with an initial V$_{bulk}$ = 0.35 km s$^{-1}$ and Q=1x10$^{26}$ H$_2$O s$^{-1}$) is simulated for a possible stealth plume under the current detection limits, however, this kind of small-scale plumes could be visible for the upcoming space missions. **Figure 5** shows the distributions of number density, velocity and column density of this modeled stealth plume. Obviously, it lacks the mostly distinct features, such as the bow-shape in the upper layer and the bowtie region in the lateral sides, seen in case 1-6 with the relatively larger gas production rates, due to insufficient gas collisions. This can be attributed to the cap effect of canopy shock layer on the plume structure as discussed in Berg et al. (2016).

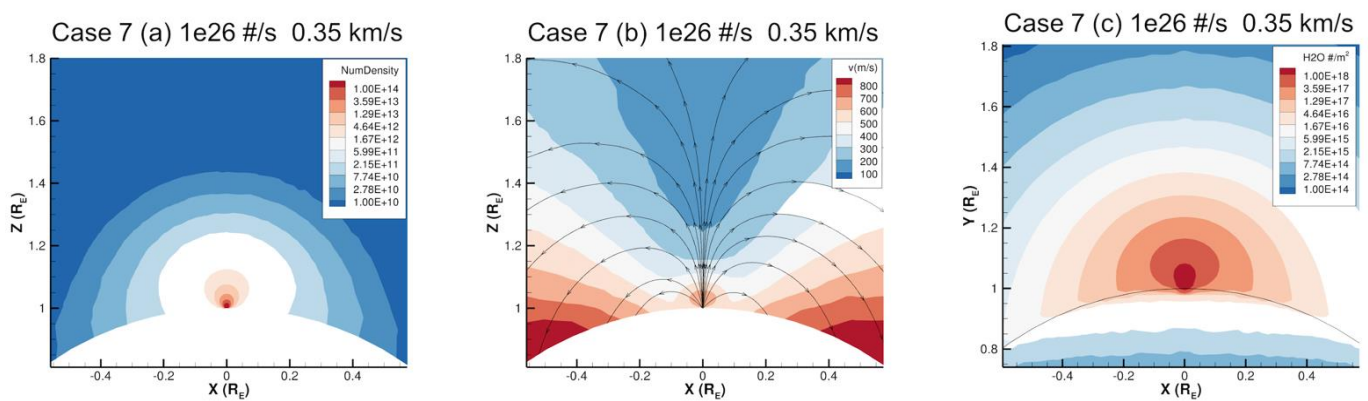

**Figure 5.** (a)Number density distribution (shown in contours with a color scale in a unit of # m$^{-3}$). (b)Gas velocity distribution (shown in contour with a color scale in a unit of m s$^{-1}$). Gas flow streamlines are presented too. (c)Column density integrated along a line of sight of 90 degrees from Europa's pole (i.e. perpendicular to the paper). The contour is shown in a color scale in a unit of # m$^{-2}$.

Both axes are arbitrary and presented in the unit of Europa radius.

The observed plume characteristics such as the gas temperature, gas and dust velocity, dust-to-gas ratio and particulate size distribution are, in fact, closely related to the undersurface vent conditions (Schmidt et al. 2008). To account for the observed dust size and speed distribution of Enceladus' plumes, Schmidt et al. (2008) proposed that dust grains are produced from nucleation inside the subsurface vent with water vapor sourced from the interface of liquid water, and wall collisions in the channel determine the grain ejection velocity. In this way, the smaller dust (sub-micron sized) would have an initial velocity exiting from the vent closer to the plume gas velocity. In addition, the width and length of subsurface channel can control the initial Mach number of the gas plumes and the critical grain size, and the measured gas flux and its dust-to-gas ratio can tell the gas temperature in the subsurface liquid water reservoir. Therefore, our DSMC model simulations, in contrast to the purely ballistic motion consideration, will serve as a useful tool to derive plume properties near surface and its undersurface vent condition.

Eruptive emplacements accompanied by cryovolcansim could explain the dark-feature material around the linea, lenticulae and ridges (Fagents et al. 2000; Fagent 2003; Quick et al. 2013). Trumbo et al. (2019) also found that the irradiated NaCl existed in surface ice of Europa's leading hemisphere, and its distribution correlated with the disrupted chaos region that is consistent with an interior source from subsurface ocean. The plume emplacements, carrying with water, salts and organic compounds from the subsurface ocean, play an important role not only on modifying Europa's surface composition but also on shaping its geological features. It could alter the surface characteristics such as albedo, color pattern, ice structure and crystallinity (e.g. Hand and Carlson, 2015; Schenk, 2020; Ligier et al. 2016; Berdis et al. 2020; Belgacem et al. 2020). For example, some bright areas in the leading hemisphere showing a narrow forward scattering behavior could be indicative of fresh frost deposit due to plume activity (Belgacem et al. 2020). In addition, while some regions with relatively higher abundance of crystalline water ice on Europa's surface could be attributed to a warmer formation environment like thermal convection in the ice shell, water outgassing emplacements could account for more amorphous ice found in the polar region (Hansen & McCord, 2004; Ligier et al. 2016). Schenk (2020) examined Europa surface images collected by Voyager, Galileo and New Horizons missions, and found no obvious surface pattern change during ~ 30 years. One of the possibilities, though relatively slim, is that the plume activity and deposit process have reached a steady state. Our simulations show nearly semi-hemisphere depositions from the large-scale plumes (i.e. with the initial gas bulk velocity in the order of 1.0 km s$^{-1}$), and therefore, the persistence of large outgassing activities may also account for Europa's young surface and hemispheric color dichotomy. In brief, the plume source region, plume dynamics and emplacements leading to the surface features are closely interlinked. Once information of one or another is brought to light by future observations, it can provide the constraints to the others. For example, knowing the exact source region coupled with the fresh deposit pattern can help constrain the gas emission velocity.

## 5. Summary

In this work, we develop a 3D DSMC model to examine the physical processes and structures of Europa's hypothetic water plumes with a parametric study regarding the total gas production rate and initial gas bulk velocity. Plume eruption is assumed with supersonic expansion from a source region 20 km in radius, which is a reasonable approach to study the large-scale structure of outgassing plumes on Europa using DSMC, though the subsurface vent condition is not well taken care of in our model. The plume characteristics (i.e. the gas number density, gas velocity and temperature distribution) derived from our model can be coupled with the radiative transfer process for the multiple-wavelength remote sensing to improve scientific interpretations of the observed spectra and images. The density profiles of the simulated plumes at different altitudes are also provided for comparisons with future measurements.

Our major findings are:

- The gas kinetic temperature rapidly drops to below 100 K during initial expansion in the low altitude region (i.e. < 100 km) due to adiabatic cooling.

- Mostly, the central parts of the plumes with large production rates of $10^{29}$ s$^{-1}$ and $10^{30}$ s$^{-1}$ are in thermal equilibrium and nearly continuum conditions.

- The gas acceleration near the surface (within an altitude of ~100 km) is a combined effect of gas collision and adiabatic cooling. At a higher altitude, the plume velocity distribution is governed by Europa's large gravity. The plume simulated from the DSMC modeling (i.e. gas collision considered) is, therefore, generally much larger than the one computed from purely ballistic motion. For example, while the purely ballistic motion of the initial velocity of 0.75 km s$^{-1}$ can arrive at a maximum altitude of only ~210 km, the similar altitude can be reached by the DSMC modeled plume of the initial velocity of 0.35 km s$^{-1}$.

- The projection effect could be significant, and so, it would be not an easy task to trace back a plume to its exact source region on Europa's surface from the observed 2D images/maps without further supporting information.

Our model inputs of high gas production rate and large gas bulk velocity producing a large-scale plume, which may be detectable by the space and ground-based telescopes, will cause semi-global deposition of Europa. While the modeling of plume dust deposition is beyond the scope of this paper, it will be studied to investigate the possible surface changes accordingly. The DSMC modeling results will be served as the background gas collision for the dust trajectory integration with a set of dust size and speed distributions. Knowledge of the magnitude, occurrence frequency and source location of the outgassing activity through the follow-up monitoring programs will advance our understanding of Europa's surface history shaped by plume deposition and erosion due to space weathering. Finally, the measurements of chemical composition of Europa's plume material can shed light on its subsurface ocean environment. The yields from the future Juno, JUICE and Europa Clipper space missions are highly anticipated to explore this icy ocean word with high potential habitability.

**Supplementary Materials:** None

**Author Contributions:** Conceptualization, Wei-Ling Tseng, Ian-Lin Lai, Wing-Huen Ip and Hsiang-Wen Hsu; Methodology, Ian-Lin Lai and Jong-Shinn Wu; Software, Ian-Lin Lai and Jong-Shinn Wu; Writing – original draft, Wei-Ling Tseng; Writing – review & editing, Wei-Ling Tseng, Ian-Lin Lai, Wing-Huen Ip and Hsiang-Wen Hsu. All authors have read and agreed to the published version of the manuscript.

**Funding:** This work is partially supported by MOST 108-2112-M-003-002 (Physics) and MOST 110-2112-M-008-003. It is also partially funded by the Einstein young scholar fellowship program (MOST 109-2636-M-003-001) and the MOE Yushan young scholar program.

**Data Availability Statement:** Not applicable.

**Conflicts of Interest:** The authors declare no conflict of interest.

**References**
1. Anderson, J. D., Schubert, G., Jacobson, R. A. et al., 1998, "Europa's Differentiated Internal Structure: Inferences from Four Galileo Encounters", Science, 281
2. Arnold, H., Liuzzo, L., Simon, S. et al., 2019, "Plasma Interaction Signatures of Plumes at Europa ", Journal of Geophysical Research: Space Physics, 125, e2019JA027346.
3. Belgacem, I., Schmidt, F., Jonniaux, G. et al., 2020, "Regional study of Europa's photometry", Icarus 338, 113525
4. Berdis, J. R., Gudipati, M. S., Murphy, J. R., et al., 2020, "Europa's surface water ice crystallinity: Discrepancy between observations and thermophysical and particle flux modeling", Icarus 341, 113660
5. Berg, J. J., Goldstein, D. B., Vargheseet, P. L. et al., 2016, "DSMC simulation of Europa water vapor plumes", Icarus 277, 370–380
6. Bird, G.A., 1994, "Molecular Gas Dynamics and the Direction Simulation of Gas Flows.", Clarendon Press, Oxford, UK.
7. Carr, M. H., Belton, M. J., Chapman, C. R. et al., 1998, "Evidence for a subsurface ocean on Europa", Nature, 391, 363
8. Cave, H. M., Tseng, K.-C., Wu, J.-S., et al., 2008, "Implementation of unsteady sampling procedures for the parallel direct simulation Monte Carlo method", J. Comput. Phys., 227, 6249
9. Cercignani, C, 1998, "The Boltzmann Equation and Its Application", Springer, New York
10. Dong, Y., Hill, T. W., Teolis, B. D. et al. 2011, "The water vapor plumes of Enceladus", J. Geophys. Res., 116, A10204
11. Fagents, S. A., 2003, "Considerations for effusive cryovolcanism on Europa: The post-Galileo perspective", JOURNAL OF GEOPHYSICAL RESEARCH, VOL.108, NO. E12, 5139
12. Fagents, S. A., Greeley, R., Sullivan, R. J. et al., 2000, "Cryomagmatic mechanisms for the formation of Rhadamanthys Linea, triple band margins, and other low albedo features on Europa", Icarus, 144, 54– 88
13. Figueredo, P. H., & Greeley, R., 2004, Resurfacing history of Europa from pole-to-pole geological mapping. Icarus, 167(2), 287–312
14. Finklenburg, S., Thomas, N., Su, C. -C., et al., 2014, "The spatial distribution of water in the inner coma of Comet 9P/Tempel 1: Comparison between models and observations", Icarus, Volume 236, p. 9-23.
15. Giono, G., Roth, L., Ivchenko, N., et al., 2020, "An Analysis of the Statistics and Systematics of Limb Anomaly Detections in HST/STIS Transit Images of Europa", AJ, 159, 155
16. Goldstein, D. B., Hedman, M., Manga, M. et al., 2018, "Enceladus plume dynamics: from surface to space. In Enceladus and the Icy Moons of Saturn "(Schenk et al., eds), pp. 175-194. Univ. of Arizona, Tucson
17. Greenberg, R., Geissler, P., Hoppa, G., & Tufts, B. R. et al., 2002, RvGeo, 40, 1004
18. Hand, K. P., Carlson, R. W., 2015, "Europa's surface color suggests an ocean rich with sodium chloride", GRL 42, 3174
19. Hansen, C. J., Esposito, L. W., Stewart, A. I. F. et al. 2008, "Water vapour jets inside the plume of gas leaving Enceladus", Nature, 456, 477–479,
20. Hansen, C. J., Esposito, L. W., Hendrix, A. R. et al., 2011, "The composition and structure of the Enceladus plume", Geophys. Res. Lett., 38, L11202, doi:10.1029/2011GL047415
21. Hansen, C. J., Esposito, L. W., Hendrix, A. R. et al., 2019, "Ultraviolet observation of Enceladus' plume in transit across Saturn, compared to Europa", Icarus 330, 256–260
22. Hansen, G. B., McCord, T. B., 2004, "Amorphous and crystalline ice on the Galilean satellites: A balance between thermal and radiolytic processes", JOURNAL OF GEOPHYSICAL RESEARCH, VOL. 109, E01012, doi:10.1029/2003JE002149
23. Jia, X., Kurth, W., Kivelson, M. et al., 2018, "Evidence of a plume on Europa from Galileo magnetic and plasma wave signatures", NatAs, 2, 459–464
24. Jia, X., Kivelson, M., Paranicas, C. et al., 2021, "Comment on "An Active Plume Eruption on Europa During Galileo Flyby E26 as Indicated by Energetic Proton Depletions" by Huybrighs et al.", GRL, 48, 6
25. Kivelson, M. G., Khurana, K. K., Russell, C. T. et al., 2000, "Galileo magnetometer measurements: a stronger case for a subsurface ocean at Europa", Science, Vol. 289, Issue 5483, pp. 1340-1343, DOI: 10.1126/science.289.5483.1340
26. Lai, I. -L., Su, C. -C., Ip, W. -H. et al., 2016, "Transport and Distribution of Hydroxyl Radicals and Oxygen Atoms from $H_2O$ Photodissociation in the Inner Coma of –Comet 67P/Churyumov-Gerasimenko", Earth Moon Planets 117: 23-39
27. Lai, I. -L., Ip, W. -H., Lee, J. -C et al., 2017, "Seasonal variations of the source regions of the dust jets of comet 67P/Churyumov-Gerasimenko", EPSC, Vol. 11, EPSC2017-784
28. Lai, I. -L., Rubin, M., Wu, J. -S. et al., 2019, "DSMC Simulation of Europa's Gas Plume", EPSC, Vol. 13, EPSC-DPS2019-575-1
29. Liao, Y., Marschall, R. Su, C. -C. et al., 2018, "Water vapor deposition from the inner gas coma onto the nucleus of Comet 67P/Churyumov-Gerasimenko", Planetary and Space Science, 157, 1-9
30. Ligier, N., Poulet, F., Carter, J. et al., 2016, "VLT/SINFONI OBSERVATIONS OF EUROPA: NEW INSIGHTS INTO THE SURFACE COMPOSITION", The Astronomical Journal, 151:163 (16pp)
31. Lo, M.-C., Su, C.-C., Wu, J.-S. et al., 2015, "Modelling Rarefied Hypersonic Reactive Flows Using the Direct Simulation Monte Carlo Method", Comput. Phys. Commun., 18, 1095
32. Marschall, R., Su, C. -C., Liao, Y. et al., 2016, "Modelling observations of the inner gas and dust coma of comet 67P/Churyumov-Gerasimenko usingROSINA/COPS and OSIRIS data: First results", A&A 589, A90
33. Paganini, L., Villanueva, G. L., Roth, L. et al., 2020, "A measurement of water vapour amid a largely quiescent environment on Europa", NatAs, 4, 266–272
34. Perry, M. E., Teolis, B. D., Grimes, J. et al., 2016, "DIRECT MEASUREMENT OF THE VELOCITY OF THE ENCELADUS VAPOR PLUMES", 47th Lunar and Planetary Science Conference

35. Porco, C., Helfenstein, P., Thomas, P. C. et al., 2014, "How the geysers,tidal stresses,and thermal emission across the south polarterrain of Enceladus are related", Astron J148, doi:10.1088/0004-6256/148/3/45
36. Quick, L. C., Barnouin, O. S., Patterson, G. W. et al., 2013, "Constraints on the detection of cryovolcanic plumes on Europa", Planetary and Space Science, 86, 1–9
37. Quick, L., Hedman, M., 2020, "Characterizing deposits emplaced by cryovolcanic plumes on Europa", Icarus 343:113667
38. Roth, L. J., Saur, J., Retherford, K. D. et al., 2014, "*Transient Water Vapor at Europa's South Pole*", Science, 343, 171–174
39. Roth, L., Retherford, K. D., Saur, J. et al., 2014b, "Orbital apocenter is not a sufficient condition for HST/STIS detection of Europa's water vapor aurora". Proceedings of the National Academy of Sciences, 111(48), E5123–E5132.
40. Roth, L. J., Saur, J., Retherford, K. D. et al., 2016, "Europas far ultraviolet oxygen aurora from a comprehensive set of HST observations", Journal of Geophysical Research: Space Physics, 121, 2143–2170
41. Schenk, P. M., 2020, "The Search for Europa's Plumes: No Surface Patterns or Changes 1979–2007?", The Astrophysical Journal Letters, 892: L12 (7pp)
42. Schmidt, J., Brilliantov, N., Spahn, F., et al., 2008, "Slow dust in Enceladus' plume from condensation and wall collisions in Tiger Stripe fractures", Nature 451, 685–688
43. Southworth, B. S., Kempf, S., Schmidt, J. et al., 2015, "Modeling Europa's dust plumes", Geophys. Res. Lett., 42, 10,541–10,548, doi:10.1002/2015GL066502.
44. Sparks, W. B., Hand, K. P., McGrath, M. A. et al., 2016, "Probing for Evidence of Plumes on Europa with HST/STIS", ApJ, 829, 121
45. Sparks, W. B., Schmidt, B. E., McGrath, M. et al., 2017, "Active Cryovolcanism on Europa?", ApJL, 839, L18
46. Sparks, W. B., Richter, M., deWitt, C. et al., 2019, "A Search for Water Vapor Plumes on Europa using SOFIA", ApJL, 871, L5
47. Spencer, J.R., Pearl, J.C., Segura, M. et al., 2006, "Cassini encounters Enceladus: Background and the discovery of a south polar hot spot", Science 311, 1401–1405
48. Spitale, JN., Porco, CC., 2007, "Association of the jets of Enceladus with the warmest regions on its south-polar fractures.", Nature, 449(7163):695-697
49. Su, C. -C., Tseng, K. -C., Cave, H. M., Lian, Y. -Y., Kuo, T. -C., Jermy, M. C., and Wu, J. -S. et al., 2010, "Implementation of a Transient Adaptive Sub-Cell Module for the Parallel DSMC Code Using Unstructured Grids," Computers & Fluids, Vol. 39, pp. 1136-1145
50. Tenishev, V., Combi, M. R., Rubin, M. et al., 2011, "NUMERICAL SIMULATION OF DUST IN A COMETARY COMA: APPLICATION TO COMET 67P/CHURYUMOV-GERASIMENKO", The Astrophysical Journal, Volume 732, Number 2
51. Teolis, B. D., Wyrick, D. Y., Bouquet, A. et al., 2017, "Plume and surface feature structure and compositional effects on Europa's global exosphere: Preliminary Europa mission predictions", Icarus, 284, 18-29
52. Trumbo, S. K., Brown, M. E., Hand, K. P. et al., 2019, "H2O2 within Chaos Terrain on Europa's Leading Hemisphere", The Astronomical Journal, 158:127
53. Vorburger, A., & Wurz, P., 2021, "Modeling of possible plume mechanisms on Europa" Journal of Geophysical Research: Space Physics, 126, e2021JA029690
54. Wu, J. –S., Tseng, K. –C., Wu, F. –Y. et al., 2004, "Parallel three-dimensional DSMC method using mesh refinement and variable time-step scheme.", Comput. Phys. Commun. 162, 166–187
55. Yeoh, S. K., Chapman, T. A., Goldstein, D. B. et al., 2015, "On understanding the physics of the Enceladus south polar plume via numerical simulation", Icarus 253:205
56. Yeoh, S. K., Li, Z., Goldstein, D. B. et al., 2017, "Constraining the Enceladus plume using numerical simulation and Cassini data", Icarus 281:357
57. Zakharov, V. V., Rodionovc, A. V., Lukianov, G. A. et al., 2009, "Monte-Carlo and multifluid modelling of the circumnuclear dust coma II. Aspherical-homogeneous, and spherical-inhomogeneous nuclei", Icarus, Volume 201, Issue 1, p. 358-380
58. Zhang, J., Goldstein, D.B., Varghese, P.L., et al., 2004. Numerical modeling of ionian volcanic plumes with entrained particulates. Icarus 172, 479–502